\documentclass[11pt,twoside]{article}
\usepackage{./asp2014}

\aspSuppressVolSlug
\resetcounters

\begin{document}

\title{VALD3: current developments}
\author{Pakhomov Yu.$^1$, Piskunov N.$^2$ and Ryabchikova T.$^1$
\affil{$^1$Institute of Astronomy, Russian Academy of Sciences, Moscow, Russia; \email{pakhomov@inasan.ru}}
\affil{$^2$Department of  Physics and Astronomy, Division of Astronomy and Space Physics, Uppsala University, Uppsala, Sweden; \email{nikolai.piskunov@physics.uu.se}}
}

\paperauthor{Pakhomov Yu.}{pakhomov@inasan.ru}{}{Institute of Astronomy, Russian Academy of Sciences}{}{Moscow}{}{117019}{Russia}
\paperauthor{Piskunov N.}{nikolai.piskunov@physics.uu.se}{}{Uppsala University}{Department of  Physics and Astronomy}{Uppsala}{}{751 20}{Sweden}
\paperauthor{Ryabchikova T.}{ryabchik@inasan.ru}{}{Institute of Astronomy, Russian Academy of Sciences}{}{Moscow}{}{117019}{Russia}

\begin{abstract}
Today Vienna Atomic Line Database (VALD) is one of main databases of atomic and molecular parameters required for stellar spectra analysis. We present the new features that recently appeared in the VALD3 release, including the effects of isotopic composition and hyperfine splitting. The latest version of VALD contains parameters for several isotopes of Li, Ca, Ti, Cu, Ba, Eu, and hyperfine splitting of 35 isotopes from Li to Eu. 
\end{abstract}

\section{Introduction}
Vienna Atomic Line Database (VALD) was created in 1992 by the astrophysicists and atomic physicists from Vienna University, Helsinki University, University of Michigan and from the Institute of Astronomy RAS (INASAN). It contains atomic parameters and parameters of spectral line required for stellar spectra analysis. Since then the database went though significant evolution and expansion. Today the data collection is combined with extensive bibliography and flexible extraction tools. For each radiative transition VALD\ provides species name, wavelength, energy, quantum number J and Land\'e-factor of the lower and upper levels, radiative, Stark and Van der Waals damping factors, full electronic description and term designation of both levels. The first release was publicly announced in 1995 \citep{vald95}, the second release, VALD2, in 1999 \citep{vald99a, vald99b}, and the latest third release, VALD3 with parameters of molecular lines, extended atomic data and updated tools, in 2015 \citep{vald3}. At present VALD offers parameters for about 1.2 million lines with accurate wavelengths in spectral range from 20~\AA\ to 200 microns and for more than 250 million lines with predicted line parameters. (Fig.~\ref{elem}). Over 2000 users from nearly 60 countries are registered in VALD.

\articlefigure{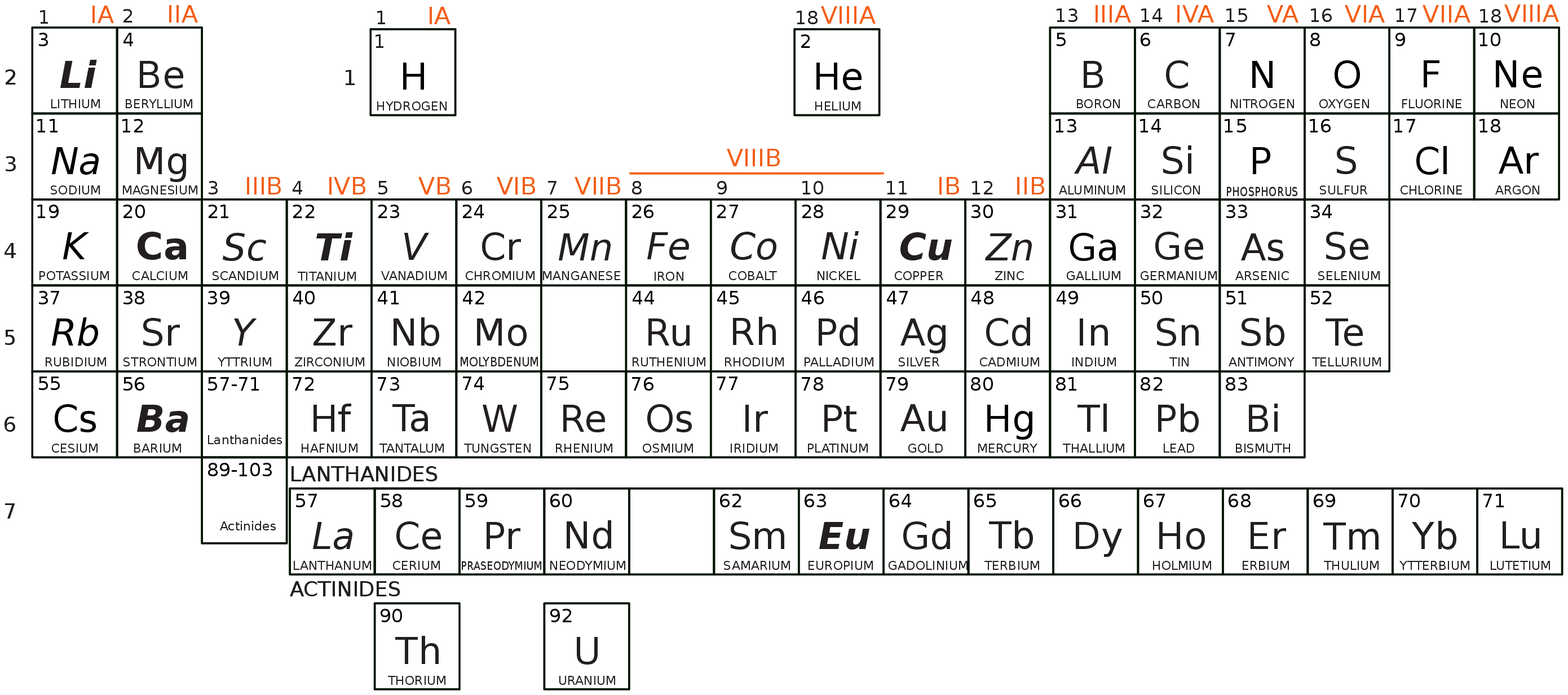}{elem}{The elements which spectral lines parameters are presented in VALD. Elements with available data for isotopic shifts are marked by bold face, and with data for hyperfine splitting are shown in italic.}

VALD is a constantly evolving database. Newly published data is continuously added allowing for improved reproduction of stellar spectra. However, there are few atomic effects, which should be included for fine analysis of high resolution spectra. The most important are isotopic composition of the elements (IS) and hyperfine splitting of atomic levels (HFS). For example, a ratio of abundances [Eu/Ba] is an important indicator for studying chemical evolution of the Galaxy because Eu is a pure r-process element while Ba is an element of s-process. Both elements have several isotopes and significant hyperfine splitting (Fig~\ref{profiles}a,b). To study the ratio [Eu/Ba] we need to know accurate abundances [Eu/H] and [Ba/H], and these effects must be taken into account. Today VALD is able to provide the IS and HFS data to its users. This article presents a description of the new functionality of VALD3.

\section{Isotopic composition}

The majority of chemical elements in the Universe is a mixture of several stable isotopes, which differ by the number of neutrons and hence, by the nuclear mass. Usually one of these isotopes dominates, and the majority of atomic databases including previous versions of VALD provide information a mixture of these isotopes. However for elements with more than one isotope with comparable abundances the spectra differ slightly due to small shifts of energy levels. This effect is observable with high resolution spectroscopy. Calculation of the isotopic shifts of energy levels is a difficult task so they are often measured in laboratory experiments. In this case, isotopic shifts may be described by two ways. The first one is the shifts of both lower and upper energy levels. The second one is the shifts of the wavelength for the same line corresponding to different isotopes. Both ways are suitable for spectrum synthesis, however, the first way is more complete. VALD3 stores energy level shift when available and falls back to wavelength shifts otherwise. 

Functional extension of VALD3 for isotopic data does not require an addition of new parameters. Each isotope is included as one of species similar to any ion, so this does not affect the speed of the query execution. 
Isotopic scaling in VALD corresponds to normal isotopic composition from \cite{isotope}. Oscillator strength of spectral line of the particular isotope is reduced by \textrm{log}~$(N_{isotope}/N_{element})$ where $N_{isotope}/N_{element}$ is the relative abundance of the isotope. Optionally, the user can extract data where all isotopes have the same total value of log~$gf$ for analysis of the stellar isotopic composition.
	
At this time VALD3 has isotopic data for Li, Ca, Ti, Cu, Ba, and Eu.

\section{Hyperfine splitting}
\label{hfs}

The effect of hyperfine splitting of the energy levels of an atom is a consequence of the interaction of the magnetic field created by the electron shell with the magnetic moment of the nucleus. In terms of quantum numbers an interaction between total moment $J$ and nuclear moment $I$ creates HFS moment $F$ which may be equal $|J-I| .. |J+I|$. As a result, the initial energy level $J$ splits into separate levels and spectral line splits into components. Selection rules between the splitted levels are $\Delta\,F=0,\pm 1$ for dipole transitions, and also $\pm 2$ for quadrupole transitions, but the transition $F_l=0 \rightarrow F_u=0$ (indices show lower and upper levels) is forbidden. The shift of the splitted levels relative to the initial energy is defined by the quantum numbers and HFS constants $A$ and $B$:
\begin{small}
$$
\Delta E = \frac{1}{2}AK + B\frac{(3/4)K(K+1)-J(J+1)I(I+1)}{2I(2I-1)J(2J-1)}
$$
\end{small}
where
\begin{small}
$$
K=F(F+1)-J(J+1)-I(I+1)
$$
\end{small}
The relative intensity of the components is calculated using 6j-symbols:
\begin{small}
$$
I(F_l\rightarrow F_u)=\frac{(2F_l+1)(2F_u+1)}{2I+1}\left\{\begin{array}{ccc}
J_l & F_l & I \\
F_u & J_u & 1~|~2 \end{array}\right\}^2
$$
\end{small}
where value 1 is used for dipole transition and 2 for quadrupole transition. The oscillator strength of particular component formed by transition $F_l\rightarrow F_u$ is reduced by log~$I(F_l\rightarrow F_u)$. The effect of hyperfine splitting on line intensity depends on the distribution of HFS components inside the line profile and its intensities. It may be significant (Fig.~\ref{profiles}c) and marginal (Fig.~\ref{profiles}d) even for the lines of the same element. 

\articlefigure{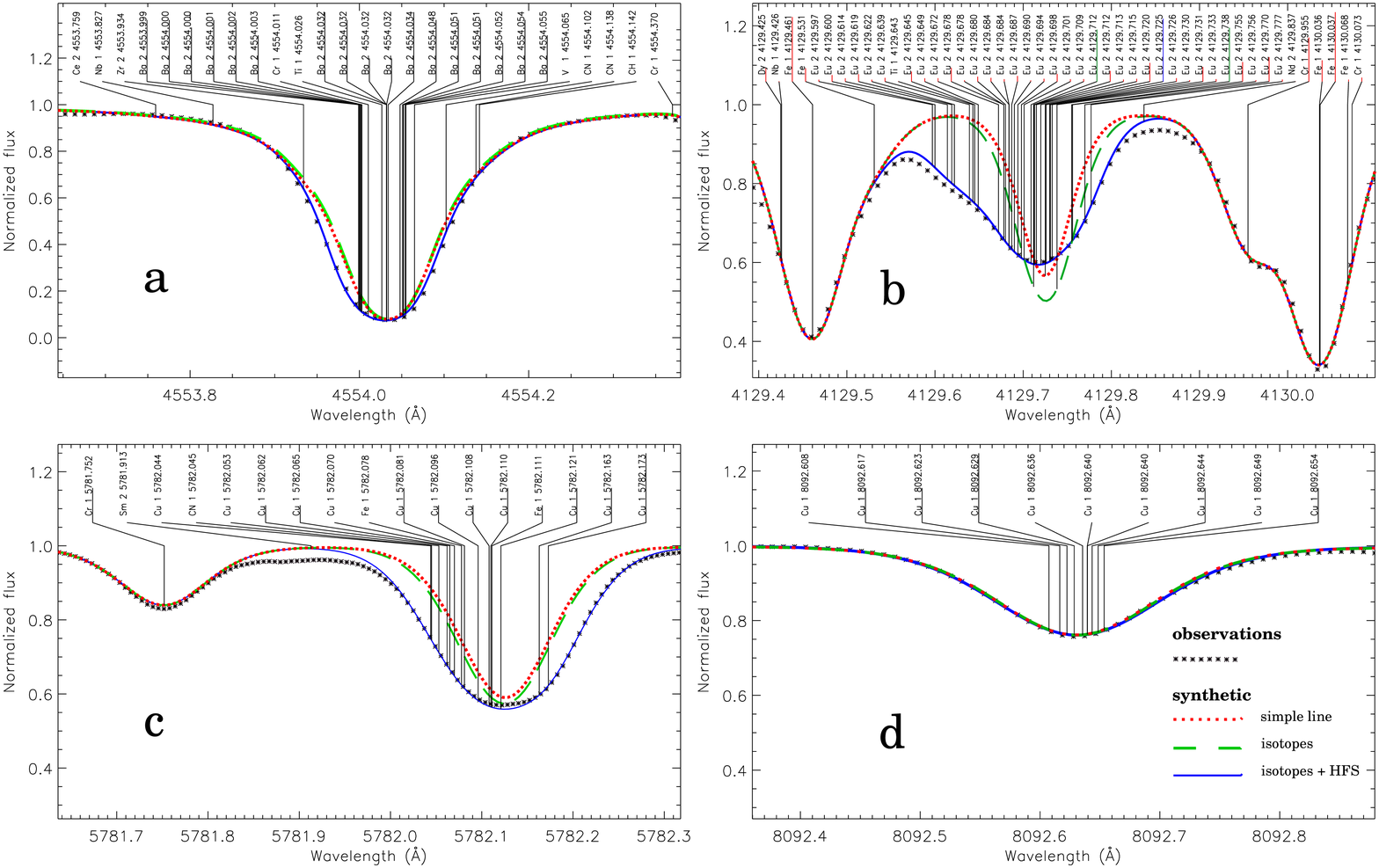}{profiles}{Examples of spectral lines profiles calculated: for a single line (accounting only for the main isotope), accounting for isotopic composition, and accounting for both the effects of isotopic composition and hyperfine splitting.}

HFS calculation is optional for VALD3 query "select stellar" and is produced on-the-fly during the line strength assessment. Unlike isotopic effects the HFS calculation requires new parameters: magnetic nuclear moment $I$, HFS constants $A$ and $B$, which are usually derived from experiments. The most recent HFS data is collected by R.~L.~Kurucz in his web site (http://kurucz.harvard.edu/atoms.html). Inside VALD3 the new option is based on modification of the Fortran program and SQLite database, which contains species (element name, element number, ion, species ID), the energy of levels $E$, the total moment $J$, the magnetic moment $I$, the HFS constants $A$ and $B$. Matching of the levels from VALD3 and SQLite is made by values of $E$ and $J$. The calculation begins if HFS constants are present for both lower and upper levels in the SQLite database. SQLite uses indexing by species ID and $E$ to accelerate the queries. Typical speed of HFS calculations is about 1800~lines/sec for CPU Intel Core i7-2600 3.4~GHz. It is a small addition to execution time of the typical VALD3 query.

Now VALD3 has the following species with HFS data: {\small $^{6}$Li~I, $^{7}$Li~I, $^{23}$Na~I, $^{27}$Al~I, $^{27}$Al~II, $^{39}$K~I, $^{40}$K~I, $^{41}$K~I, $^{45}$Sc~I, $^{45}$Sc~II, $^{47}$Ti~I, $^{49}$Ti~I, $^{47}$Ti~II, $^{49}$Ti~II, $^{50}$V~I, $^{51}$V~I, $^{51}$V~II, $^{55}$Mn~I, $^{55}$Mn~II, $^{57}$Fe~I, $^{59}$Co~I, $^{59}$Co~II, $^{61}$Ni~I, $^{63}$Cu~I, $^{65}$Cu~I, $^{67}$Zn~I, $^{67}$Zn~II, $^{85}$Rb~I, $^{87}$Rb~I, $^{89}$Y~II, $^{135}$Ba~II, $^{137}$Ba~II, $^{139}$La~II, $^{151}$Eu~II, $^{153}$Eu~II}.

\begin{table}
\vskip -10pt
\caption{Selection from VALD for the range of 6707.5 - 6708.0~\AA\ with lithium resonance lines.}
\label{Li}
\begin{center}
{\footnotesize
\vskip -10pt
\begin{tabular}{cccccccc}
\tableline
\noalign{\smallskip}
Species & Wavelength, \AA & $E_{low}$, eV & log~$gf$ &
Species & Wavelength, \AA & $E_{low}$, eV & log~$gf$
\\
\tableline
\noalign{\smallskip}
    CN  &       6707.5442&   0.9557&  -1.600&$^7$Li I&       6707.9064&   0.0000&  -0.842\\
$^7$Li I&       6707.7558&   0.0000&  -1.240&$^7$Li I&       6707.9078&   0.0000&  -1.541\\
$^7$Li I&       6707.7558&   0.0000&  -0.842&$^7$Li I&       6707.9185&   0.0000&  -0.842\\
$^7$Li I&       6707.7559&   0.0000&  -0.842&$^6$Li I&       6707.9192&   0.0000&  -1.951\\
$^7$Li I&       6707.7679&   0.0000&  -1.541&$^6$Li I&       6707.9192&   0.0000&  -1.854\\
$^7$Li I&       6707.7680&   0.0000&  -0.842&$^7$Li I&       6707.9199&   0.0000&  -0.842\\
$^7$Li I&       6707.7681&   0.0000&  -0.395&$^6$Li I&       6707.9226&   0.0000&  -2.854\\
    Fe I&       6707.7854&   5.6207&  -2.323&$^6$Li I&       6707.9226&   0.0000&  -1.951\\
    CN  &       6707.7994&   1.2059&  -1.973&$^6$Li I&       6707.9227&   0.0000&  -1.423\\
\tableline\
\end{tabular}
}
\end{center}
\vskip -25pt
\end{table}

Table~\ref{Li} shows the extraction from VALD3 for the solar atmosphere in the range of the resonance lithium line 6707~\AA, which is consists of 15 HFS components of two isotopes $^{6}$Li~I and $^{7}$Li~I. Previous versions of VALD included two lines of doublet Li~I. Taking into account isotopic composition and ignoring the HFS we would get only two pairs of lines.

\acknowledgements This research was supported by Basic Research Program P-7 of the Presidium of the Russian Academy of Sciences and grant of the Leading School No\,9951.2016.2.



\end{document}